\documentclass[conference]{IEEEtran}
\IEEEoverridecommandlockouts
\usepackage{amsmath,amssymb,amsfonts}
\usepackage{algorithmic}
\usepackage{graphicx}
\usepackage{textcomp}
\usepackage{xcolor}

\usepackage[sorting=none]{biblatex}

\def\BibTeX{{\rm B\kern-.05em{\sc i\kern-.025em b}\kern-.08em
    T\kern-.1667em\lower.7ex\hbox{E}\kern-.125emX}}
\begin{document}

\title{Xcrum: A Synergistic Approach Integrating Extreme Programming with Scrum \\
}

\author{
    \IEEEauthorblockN{Siavash Hosseini}
    \IEEEauthorblockA{\textit{Department of Electrical and Computer Engineering, Lakehead University,Thunder Bay, ON P7B 5E1, Canada} \\
    shossei4@lakeheadu.ca}
}

\maketitle

\thispagestyle{plain}
\pagestyle{plain}

\begin{abstract}
In today's modern world, software plays a pivotal role. Software development is a highly complex and time-consuming process, demanding multidimensional efforts. Companies continually adapt their requirements to align with the evolving environment, with a specific emphasis on rapid delivery and the acceptance of changing requirements. Traditional models, such as plan-driven development, often fall short in meeting these demands. In the realm of software development, Agile has been the focal point of global discourse for both researchers and developers. Agile development is better suited to customize and streamline the development process, offering a highly flexible, early, and rapid delivery lifecycle conducive to efficient software development. This article aims to provide an overview of two prominent Agile methodologies: Scrum and Extreme Programming (XP). It achieves this by reviewing relevant publications, analyzing their impact on software development, exploring the distinctive features of each methodology, and conducting a comparative assessment. Furthermore, the article offers personal insights and recommendations. Notably, the integration of XP practices into Scrum has given rise to a novel hybrid methodology known as "Xcrum," which retains its agility. It should be highlighted that, given this new approach's incorporation of the strengths of both methods, it holds the potential to outperform the original frameworks.

\end{abstract}


\section{Introduction}
There are different kinds of system development methods \cite{jayaratna1994understanding} namely; structured \cite{eva1994ssadm}, object- oriented \cite{beynon1998object}, waterfall \cite{cheatham1991object} etc. Agile methodology is another group of proposed methods which was introduced first time in the late 1990s. Agile methods are popular for many companies and has been used in industries for couple of years internationally \cite{charette2001decision}. Since agile is flexible, it has several approach for different projects. Agile is a method with continuous iterations and testing during the process of development. As a result of the collaboration between self-organizing, cross-functional teams, requirements and solutions will evolve \cite{collier2012agile}. Among different agile methods, scrum and XP have received the most attention \cite{cobb2015project}. These two methods can be used in different situation and have their own pros and cons. However, utilizing the two mentioned methods in wrong approach will be inefficient which lead to squandering time and resources. Scrum is a lightweight framework that helps the development team to focus on delivering the highest business value in the shortest time. In addition, scrum can be considered as the combination of incremental and iterative models since the builds are basically incremented in the terms of developed features \cite{schwaber2002agile}. 

Extreme Programming (XP) is another methods under agile methodology which was proposed by Kent Beck. XP is also a lightweight framework that is more suitable for fast-changing environment because of its flexibility \cite{newkirk2002introduction}. Scrum and XP are more likely to be used in small scale projects and organizations.Both of them called lightweight frameworks due to removal of formal activities to increase agility and simplicity. It is worth mentioning that, scrum and XP have several common and contrasting points in their methodologies. The purpose of this study is to analyze and compare them more specifically. Agile methodology emphasizes adaptability and iterative improvement, allowing for flexible adjustments and continuous learning throughout the development process \cite{hosseini2020precise,hosseini2021application,hosseini2021increasing,hosseini2022accurate,hosseini2023advanced, bahiraei2020neural,bahiraei2021neural,bahiraei2021predicting,9918222,hosseini2023advancedt}.

\section{Agile methodology}
Agile software development was initiated with the purpose of introducing incremental and iterative approach in the process of software development \cite{srivastava2017scrum}. The core  fundamentals of this method defined as The Agile Manifesto \cite{beck2001manifesto}. By using the following statements, it reveals the understanding of software development requirements:
\begin{itemize}
  \item It is more important to collaborate with people than the system process.
  \item It is crucial to deliver software product than document the process.
  \item The satisfaction of the customer is more important than the negotiation.
  \item Constantly ready to accept changes in the system, when and where they are needed.
\end{itemize}

Agile Manifesto is brief and straightforward to understand and employ but in some aspects complex to handle in projects. The manifesto is based on knowledge derived from experience of software development and management practices\cite{saleh2019comparative}.

The notion behind agile manifesto emphasise on real world software management and aim to propose new ways of thinking in the development team prospective and focus on responsibility, commitment, self-organizing and accountability that should be managed by the team instead of relying on manager or supervisor. This point of view led to invention of different frameworks such as scrum and XP. These frameworks are based on agile methods which initiated based on manifesto \cite{anguelov2019research}.

\subsection{Scrum}
Scrum is a software development and management framework, introduced in 1993 and is based on agile methodology principles \cite{sutherland2011scrum}. Nonaka and Takeuchi who first time used scrum, announced that the new framework will help the world of software development to change and being adapted for corporations working in software development field \cite{taibi2017comparing}.

Scrum enables team capability to address sophisticated challenges in development process to deliver value to businesses by focusing on creativity, productivity and collaboration \cite{Schwaber2017}. 
Agile aims to mitigate the problems in traditional models. In scrum all of the released has been scheduled based on time, customer demand and quality \cite{anwer2017comparative}. 
Scrum comes with incremental release which make the whole product in small cycles named sprint. The duration of each sprint is ranging from 2 to 4 weeks.

There is a definite link between implementation of agile and these five core values: courage, management, openness, respect and focus. The team will recognize and learn these values when they are involved into self-organization.  Scrum master, product owner and team members comprehend and execute self-organization in the whole sprint to achieve the goals \cite{merzouk2017comparative}.

Substantial events that have been defined in scrum as a life cycle of scrum in one sprint are as follows: sprint, sprint planning, sprint review, sprint retrospective and daily scrum. scrum is much flexible during a sprint by defining a specific control mechanism that allows to make limitation on work in process for development. As a result they are free to focus on specific goals in spring for increment release by taking into account the DoD (definition of done).Daily scrum is essential for developers in the team to share their opinion and suggestions and highlight their impediments. Daily scrum can be considered as an example of self-organizing team. sprint review gives an opportunity to scrum team as well as stakeholders to analyze and review the results. It is worth mentioning that sprint retrospective provide an opportunity to plan improvements for the next sprint \cite{bhavsar2020scrum}.

Scrum offers a limitation for number of team members. Collaboration is a matter of key importance in scrum whilst in large groups(more than 9 members), interaction and communication between individual is more challenging. Limited members in team reduces conflicts and increases velocity \cite{ns2000scrum}.

In 2010, a research item has been published which studied the future of scrum to show how scrum increases productivity in projects and concluded that frame works based on agile are more efficient in software projects. According to their obtained results, utilizing scrum is a crucial to improve and increase productivity \cite{cardozo2010scrum}. Harleen evaluated different methods of agile in companies working on mobile application development, it has been realized that methods like scrum and XP are more suitable for their units. As it has been proved frequently scrum will result in improvement of customer values \cite{flora2014adopting}. 

\subsection{Scrum phases}

\begin{figure}[h]
\includegraphics[width=\linewidth]{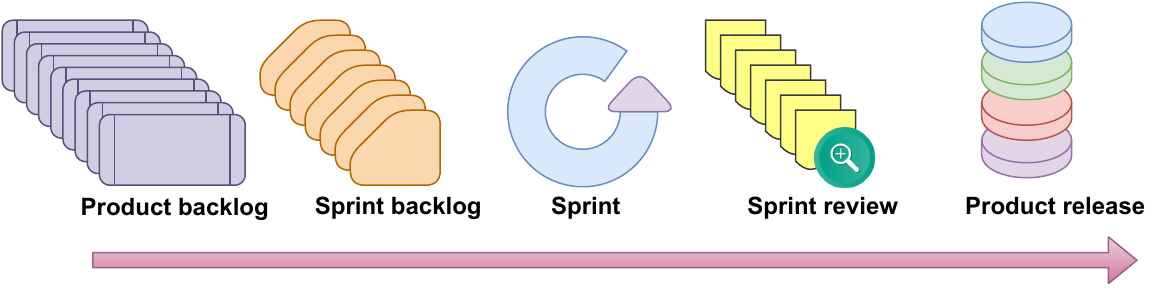}
\caption{Scrum flow}
\label{fig:scrum_flow}
\end{figure}

Scrum workflow comprises different stages. Fig \ref{fig:scrum_flow} shows scrum flow for projects. firstly, items are sent into product backlog.Then, development team send items into sprint backlog to accomplish task and meet DoD of the sprint. Following this, in spring review process, completed stories will be verified to deliver as an increment. 

Different phases in scrum can be grouped in three comprehensive stages namely: pregame, game and post game.

Pregame: By defining the purpose of the project while it is still unclear, this stage begins. Product owner is the role in scrum who is responsible to make clear this uncertainty and prepare list of prioritized functions. This list is known as product backlog. A single list also will be prepared by development team which includes timing schedule, cost, delivery date and number of releases. There are also some critical tasks in this phase such as risk assessment, development tools and required funds. 

Game: This is the stage that is related to development and called sprint in scrum. Each sprint is combination of development, review and adjustment.

Post game: After developing the negotiated features, finally it is the time for final release. In this stage, final integration testing will be carried out and user manuals and training materials are ready for finalization and release the final version.

\subsection{Sprint cycle}
Scrum works in time boxed duration called sprint which development and coding happened. During this time, team works on tasks and feature under the guidance and supervision of scrum master. Also there are some key activities in scrum namely; Sprint planning, daily scrum, sprint development, sprint review and sprint retrospective.

\subsection{Scrum roles}
In scrum 3 important roles can be defined. Scrum master, product owner and development.

Product owner: The customer's representative who is responsible for the product. The list of features will be created and prioritized by product owner for development in the shape of product backlog. He can change the position of each feature based on business needs also he is responsible for scheduling and providing financial needs. He talks with the team to make them aware about the stakeholders enthusiasm. For the sake of the job, he must have comprehensive understanding about business, engineering and market. Great communication skill is a matter of key importance to work with wide variety of stakeholders.

Scrum master: Scrum master is responsible for tracking activity of the team to make sure they are following scrum practices, roles and values. This role is totally different from traditional project manager. He protects team against circumstances and tries to remove impediments.

Team: Scrum teams are self-organizing which has 3 to 9 members. In scrum teams, they divide tasks according to their interest. All of the team members should have appropriate skills to design, develop and test the product.

Scrum is a framework under the hood of agile which is based on empiricism and these are main reasons that scrum has been more and more popular as a standard and efficient approach toward software development. The results of comprehensive investigation have demonstrated that, taking advantages of proper implementation of this framework in software development practices will improve the quality and productivity in developed features \cite{bhavsar2020scrum}.

\subsubsection{Strength of scrum}

The strength of scrum depends on these three factors: transparency, inspection and the last but not least, adaptation.
Transparency means that everything in the project that influence the process should be clear and visible to everyone in the product development. Inspection meaning is to spectate and track all of the activity and process to recognize unacceptable output. Ultimately, adaptation is the term which means adjustment when specific output is not acceptable \cite{schwaber2002agile}.

Employing scrum is pretty much easy but there are some sophistication in the term of managing. There are several characteristics that make scrum strong enough to handle massive burden of managing projects. Table \ref{table:Scrum_artifacts} shows the characteristics and artifacts of scrum.

 \begin{table}[]
 \centering
 \caption{Scrum artifacts\cite{bhavsar2020scrum}}
\label{table:Scrum_artifacts}
 
\begin{tabular}{|l|l|}
\hline
\textbf{Artifacts} & \textbf{Characteristics}     \\ \hline
Product Backlog    & Transparency, Priority Order \\ \hline
Sprint Backlog     & Work Items                   \\ \hline
Increment          & Releasable Feature           \\ \hline
\end{tabular}
\end{table}

\subsubsection{Recommendations}

In this work, it has been realized that scrum characteristics makes it properly suitable for software development process. It is recommended that organizations should pay more attention into limitations in their organization and structures.

\subsection{Extreme Programming}
Extreme Programming (XP) is one of the well-known agile methods. XP has been initiated for various associations with extraordinary size and preparations across the world \cite{saleh2017comparative}. 

XP came with a multitude number of benefits and successes. This method made great deal of interest due to its realistic way for dealing with improvement. XP has been developed by Kent Beck in 1996. Besides being lightweight and flexible, it is capable to adapt rapid changes in software development systems \cite{newkirk2002introduction}.

XP can be defined with 12 practices. These practices are considered as the foundations of XP. They are all together shape the whole framework. All of the practices are shown in table \ref{table:XP_REHEARSES}.

\begin{table}[]
\centering
 \caption{XP REHEARSES EXPLANATIONS\cite{newkirk2002introduction}.}
\label{table:XP_REHEARSES}

\begin{tabular}{|l|l|}
\hline
\multicolumn{1}{|c|}{Rehearses}                                       & \multicolumn{1}{c|}{Explanation}                                                                                                                                                                                                                                                                                                            \\ \hline
\begin{tabular}[c]{@{}l@{}}Small\\ Releases\end{tabular}              & \begin{tabular}[c]{@{}l@{}}Get the framework into generation apace. This is a key\\ factor in aiming input on the real programming\end{tabular}                                                                                                                                                                                             \\ \hline
Metaphor                                                              & \begin{tabular}[c]{@{}l@{}}See how the entire framework functions. It is similar\\ critical for the client to comprehend the representation\\ alongside the software engineers\end{tabular}                                                                                                                                                 \\ \hline
\begin{tabular}[c]{@{}l@{}}Simple\\ design\end{tabular}               & \begin{tabular}[c]{@{}l@{}}One from the central qualities is straightforwardness.\\ The framework ought to exist intended on behalf of the\\ highlights that follow executed today. Let the futurity\\ direct how the framework develops to that degree, don't\\ attempt or foresee the imminent, you’ll likely not be\\ right\end{tabular} \\ \hline
Testing                                                               & \begin{tabular}[c]{@{}l@{}}Criticism is additionally the another key quality.\\ Testing incorporates unit tests, which developers\\ compose, acknowledgment tests, which clients\\ compose. Test is the pointer of fruition\end{tabular}                                                                                                    \\ \hline
Refactor                                                              & \begin{tabular}[c]{@{}l@{}}Developers are in charge of enhancing the plan of\\ existing programming without changing its conduct.\\ Refactoring is a piece of the software engineer's regular\\ exercises\end{tabular}                                                                                                                      \\ \hline
\begin{tabular}[c]{@{}l@{}}Pair\\ Programming\end{tabular}            & \begin{tabular}[c]{@{}l@{}}Functioning with an accomplice remains a prerequisite\\ during writing the implementation code\end{tabular}                                                                                                                                                                                                      \\ \hline
\begin{tabular}[c]{@{}l@{}}Collective\\ ownership\end{tabular}        & \begin{tabular}[c]{@{}l@{}}Anybody on the group can change any piece of the\\ framework\end{tabular}                                                                                                                                                                                                                                        \\ \hline
\begin{tabular}[c]{@{}l@{}}Continuous\\ integration\end{tabular}      & \begin{tabular}[c]{@{}l@{}}Developers incorporate and assemble the product all\\ the time\end{tabular}                                                                                                                                                                                                                                      \\ \hline
\begin{tabular}[c]{@{}l@{}}Sustainable\\ working\\ hours\end{tabular} & \begin{tabular}[c]{@{}l@{}}A superior name is a work until tired. Be that as it may,\\ know about the drawbacks of working an excessive\\ number of hours numerous weeks in succession\end{tabular}                                                                                                                                         \\ \hline
On-site client                                                        & \begin{tabular}[c]{@{}l@{}}The client is in the group, accessible to response\\ inquiries in time. The client is likewise in charge of\\ composing acknowledgment tests\end{tabular}                                                                                                                                                        \\ \hline
\begin{tabular}[c]{@{}l@{}}Coding\\ standards\end{tabular}            & \begin{tabular}[c]{@{}l@{}}Correspondence is a key esteem. Receiving coding\\ models enhances correspondence in light of the fact\\ that the code is predictable from class-to-class\end{tabular}                                                                                                                                           \\ \hline
\end{tabular}
\end{table}

XP is also suitable for small and medium scale projects. XP comprises various practices and principles in disciplinary manner. XP utilizes all of the aforementioned practices in excessive way to develop high quality product. The basis of XP has more emphasise on the satisfaction of customers. Quick feedback and increment release are the key factors which aim to lessen the cost and reduce the defects \cite{beck2000extreme}.

\subsubsection{XP phases}
\begin{figure*}[h]
\includegraphics[width=\linewidth]{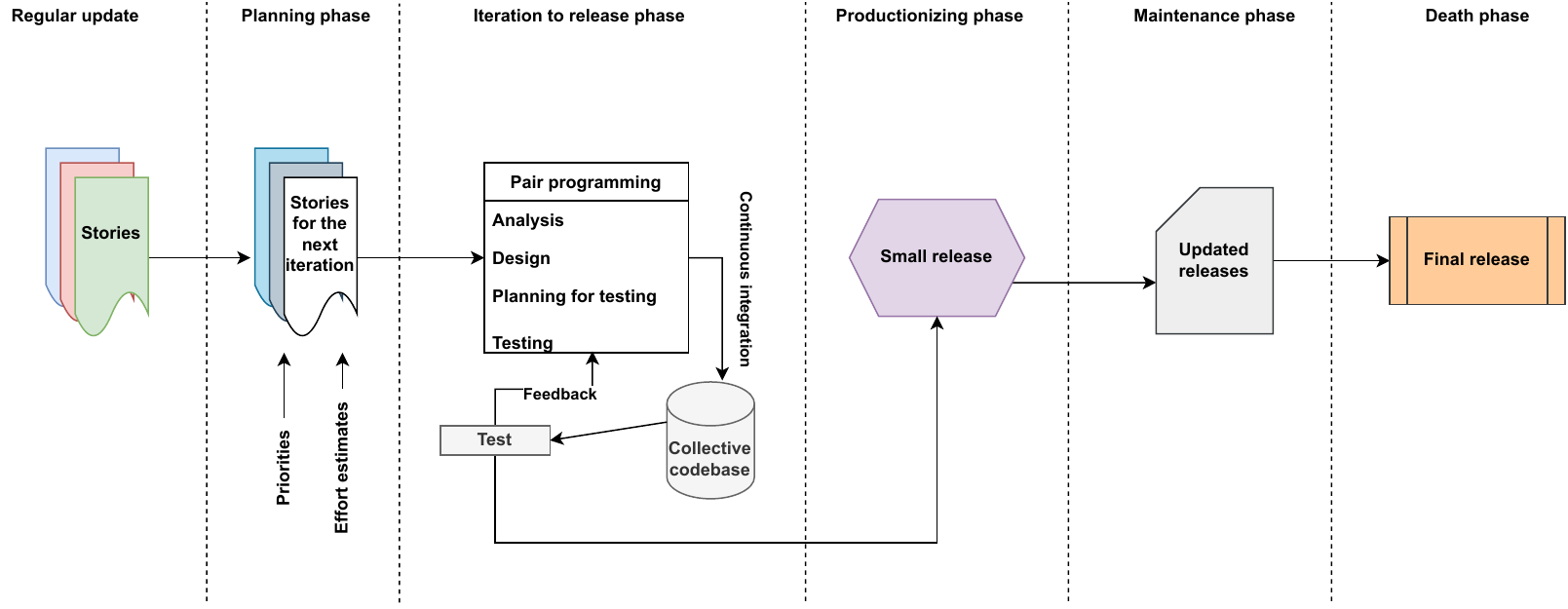}
\caption{XP phases \cite{abrahamsson2017agile}}
\label{fig:xp_phases}
\end{figure*}

Development process in XP includes 6 main phases namely: exploration, planning, iteration to release, productionizing, maintenance and death. Fig \ref{fig:xp_phases} shows different phases in XP.

Exploration phase: This is the first phase in XP life cycle. In this phase, requirements and architecture will be defined. Several meetings between customers and developers take place and customer defines stories with  explanation for each of them to be clear enough for developers.

Planning phase: Goal of this stage is to recognize what can be done until the due date that has business value and what should be done for the next iteration.

Release planning: The objective of this stage is to realize the delivery time and features that should be prepared for the due date. Customers and developers participate in meetings.

Iteration planning: Iterations commence with iteration planning. Developers should provide their activity plan for accomplishing anticipated features for the following release. In this phase developers choose tasks and give an estimation regarding the cost, effort and time which is required and at the end other programmers give their opinion to balance the load of work.

Iteration to release: This stage is combination of design, code,test and integration. This is an iteration phase and each of iterations may take 1 to 4 weeks. In the initial iteration, developers select stories that leads to creating overall figure of the system. Programmers design and write codes for selected features. After that, functional testing is applied and code is integrated. If the code does not meet the requirement, code refactoring is used. Stand up meeting take place to discuss progress of work, problems, impediments and finding solutions \cite{beck2000extreme}.

Productionizing Phase: In XP software releases in increment. Frequent releases will bread to build anticipated system gradually and in small releases. Productionizing is for checking whether the software is acceptable in the term of testing, and is it ready for releasing as a product.
Maintenance phase: In this stage, team is working on new features and software is evolving while the old version is still running. Although new functionalities are introducing, team should pay more attention to the software in production \cite{abrahamsson2017agile}.

Death phase: In this phase, all of the negotiated features have been developed and customer is totally satisfied and no stories left. Then, it is the right time to finalize the product and release the software \cite{abrahamsson2017agile}.

\subsubsection{XP values}

There are five XP values which have to be considered while XP practices are applied. These values are as follows: simplicity, communication, feedback, courage and respect \cite{abrahamsson2017agile}. 

Simplicity: emphasis is on simplicity in XP in a way that we have to keep everything in the simplest way possible. There is no complex functionality in the work until the customer request for it

Communication: Similar as scrum, XP focus on communication rather than documentation. Developers and customer communicate for finding economical and proper solution for existing problems.

Feedback: Continuous feedback and communication are the most important elements that help the team to keep the project along the line.

Courage: Due to some reasons, sometimes the team should refactor the implemented design and code. Courage means making such hard decisions for project that has not been made beforehand.

Respect: Since in XP pair programming is highly important, showing respect to other team members make it possible to implement XP practices while working on the project. Respect force the team members especially developers to deliver high quality work \cite{beck2000extreme}.

\subsection{Roles in XP}
There are seven roles for team members. Roles and their responsibilities can be found in the following:

Programmer: Writing codes is the most important part in XP which is done by programmer. XP team does not have designer and analyst and all of the tasks related to them is carried out by the programmer of the team.

Customer: Customer plays an important role during the development process. Customer is responsible for defining stories and deriving test for analyzing functionality.

Coach: Coach should be able to manage the team and also have technical skills. Good decision making ability help the coach to keep the team on right flow.

Tracker: Tracker is responsible for collecting metrics such as load factor about the project. This member of the group gather information from all of the developers and record them to have better view about spent time and time which is needed for completing the task.

Tester: Tester helps customer to write functional test and verify it. Since, unit testing has been carried out earlier by programmers, tester have less duties in XP.

Consultant: In some cases that team requires technical help, a consultan who is expert in that field can be hired and developers should have meeting to find solutions.

Big boss: This role in XP act as a coordinator which is responsible for providing w equipment and tools.

\section{Related works}
Several works have been done in this field to do a comparative study and propose a new methodology. Merzouk et al. studied comparison among XP, scrum and feature driven development to highlight pros and cons of each method. In the next step they used key specific features to suggest new model with new functionality that may be able to overcome some deficiencies in previous models \cite{merzouk2017comparative}. Herdika et al. studied similarity between practices of different agile methodology \cite{herdika2020variability}. They mentioned eight prominent similar practices among agile methods. Some theories have been proposed for scrum to help the development team to discover the user needs and organize the user stories.

Merzouk et al. conducted a research item to propose new method that improve some of deficiencies in agile methodology especially scrum and XP \cite{merzouk2018agile}. They mentioned that agile is not accurate enough to estimate demanded effort and required budget. They directed their analysis to propose new idea by combining practices which have proposed for other methods such as Crystal family. 

Anwer and co-workers, studied XP to add some new features in order to improve this method \cite{anwer2018empirical}. Although came in handy in many projects, there are still some shortages in design aspect such as poor architectural aspects. They were of the opinion that, pair programming is not a beneficial feature in all of the projects. As a result of their attitide, they came with a new method called Simplified Extreme Programming (SXP) to cover and revise XP without changing the agility.

Raza et al. investigated change management in agile method and how it can be possible to make this much flexible by using different frame works like: XP, scrum, lean, ASD etc \cite{raza2018managing}.

\section{Comparison between Scrum and XP and proposing an enhanced methodology called "Xcrum"}
Both of these frameworks are popular and highly used agile methodologies which have some similar and contrasting points.

Both models emphasise on building functional product by utilizing adaptive procedure. Incremental and iterative are two of the key features of these models however duration is different. In scrum, sprints are vary from 2 to 4 weeks but in XP, it is 1 to 3 weeks. XP offers twelve principles for guidance of project flow while in scrum team members are responsible to choose development practices \cite{abrahamsson2003new,abrahamsson2017agile}. The difference between scrum and XP is subtle but may have huge effect on the team activities. They are almost alight with each other. In my opinion there are four major differences in this case. 

\begin{itemize}
  \item People in scrum usually work in sprints which are 2 weeks or 4 weeks but in XP teams work in iterations that are one or two weeks. 
  \item In scrum we cannot allow any changes during sprints. After sprint planning meeting and defining some of the items in product backlog to be delivered, every thing remain unchanged till the end of iteration. XP is more flexible to accept sudden changes during iterations. If the team has not started working on a specific feature, new feature with equal size can be brought into iteration instead of working on unstarted one.
  \item team members in XP work in strict priority  order. Features to be developed should prioritize by the customer (equivalent with product owner in scrum) and the XP team should work on them is specified order.In sharp contrast, scrum product owner prioritizes the product backlog but in the end team will determine the sequence in which they want to develop backlog items. 
  \item  There aren't any engineering practices in scrum whilst XP prescribe some. XP engineering practices are really popular especially test-driven development, automated testing, pair programming, simple design and refactoring. These practices are really important but I believe it is not fair to mandate them. It is much better for teams to discover them by their own. 
\end{itemize}
The aforementioned descriptions are the main differences between scrum and XP. Some experts in this field recommend that it is better to start with scrum methodology and then elaborate it to XP and invent their own method which has better efficiency. XP practices are excellent but in my point of view it is far better if teams discover them instead of applying them in mandatory manner.

Some experts are of the opinion that, professional scrum team will use and apply XP practices. This kind of attitude will lead to emerging new combined methodologies.Scrum teams who are commitment to accomplish all aspects of their tasks will discover that they need XP practices in the case they want to move forward by all of their capability in the upcoming sprints. In the following some of the XP practices that can be used in scrum for strengthening this methodology are explained:
\begin{itemize}
\item In XP there is a role called XP coach pretty much similar to scrum master in scrum. XP coach is responsible for leading the team to apply XP values during iterations and unlike the scrum master, also advice the development team on technical practices. 

In scrum, it has been defined that scrum master does not need technical skills. Scrum master with technical skills and background definitely has more privileges for the team as long as development team do not expect that person deal with all of the problems and propose solution. I believe, it is more logical for teams who wants to use XP practices to hire a XP coach outside their team to assist development of technical aspects till the time they get acquainted with.

\item Scrum in general does not define the way teams are composed. It takes a look at every single member in team as a development team member. Expectation in scrum is about being cross-functional in a way that, the team can deliver releasable increment in each sprint.

\item Similar to scrum in XP all skills are required for team to work on stories and turn them into releasable product. Technical writers and interaction designers are in XP team. If the software needs someone with particular skill, it is crucial to add that role and member to the team. As a result of XP values(customer involvement), user is part of the team.

\item A sprint in scrum is a planning cadance instead of being release cadence. Similar to scrum, it is the same story for XP. Iterations in XP is a planning cadence rather than being release cadence \cite{WinNT}. Unlike scrum that it is possible to define duration of sprints between 2 weeks or 4 weeks, there is no permission to increase iterations more than one week in XP. 

There is an obligation in XP regarding duration of iterations because XP requires 40 hours a week sustainable work. If we want to apply this practice in scrum, team members must adjust the number of stories they want to work on and spend 40 hours per sprint. 

One week sprint means that, customer can receive much faster feedback from development team. In addition, it is easier to plan 1 week iteration and also it is simple to estimate future works and manage time to accomplish them. This can be considered as one of the greatest advantages of XP and this is why it has been said that XP is compatible with sudden changes.

By combining scrum and XP,  scrum Team can benefit from XP practices which means that they can select the stories that will be completed in a one week Sprint. There is dependency and meaningful connection between these stories which is really important in scrum. In sprint planning meeting team breaks down stories into tasks and divide them among development team to work on.

\item Pair programming is one of the substantial features of XP. XP team does pair programming during all of the Sprint.

Many managers belive that, pair programming will come at huge cost but it should be noted that it is only not about two people working on the same computer. Pair programming is a micro feedback loop. The more focus is about code review and revision. By taking this approach XP team would be able to find bugs and problems much more earlier as a result of collaboration and contribution of two programmer . 

Scrum Team which applies XP practices will put pair programming in the definition of "Done". Pair programming configuration may be two developers working with one machine or a programmer and a tester working together.

\item Test First Development is a valuable practice for Scrum team. Scrum team goal is to deliver high quality software. By taking advantages of test first development method, team will receive quick feedback about their work. In this case, developers have to change their code continuously and move towards a simple design. 

\item Transparency is a matter of key importance in scrum. There is no straight forward and clear approach in scrum regarding how make the documents transparent for every one. Scrum allows to utilize different tools for product backlog management. On the flip side, XP emphasise on informative work space. As a result anyone can come to working room and be aware of development progress.
\end{itemize}

By utilizing XP practices as it has been discussed in previous section, scrum methodology will be more robust compared to its original format. XP practices are really important and valuable to improve functionality in development team and by utilizing them in scrum, it would be possible to enhance efficiency in scrum. Since the practices of XP have been used in scrum, this method is not the former method any more. Then, proposing a name can be helpful to omit misconception in this case.

Combination of these two models will come with a new name called "Xcrum" which combines the name of XP and Scrum and carries both key characteristics. It is worth mentioning that by combining methods, the agility of them will not be affected and these methods are still under the hood of Agile methodology.

\section{Conclusion}
In conclusion, this study has conducted a comprehensive comparative analysis of two prominent agile methodologies, Scrum and Extreme Programming (XP). The examination has illuminated the distinctions between these methodologies, offering a profound insight into their individual strengths. With this enhanced understanding and the recognition of the advantages inherent in each approach, there is a compelling case for their integration into a unified method or for the adoption of Scrum enriched with XP practices.

At this stage, there is a strong conviction that the incorporation of XP practices can significantly enhance the development process, surpassing the effectiveness of implementing either of these methodologies in isolation. The proposed amalgamation, termed "Xcrum," leverages the benefits of XP practices while preserving the core tenets of both Scrum and XP. This synthesis remains firmly rooted within the agile framework, ensuring that the fundamental principles of both models remain intact. Consequently, "Xcrum" emerges as a promising agile subbranch that offers a harmonious blend of Scrum and XP practices to optimize software development processes.

\newpage
\AtNextBibliography{\footnotesize} 
\printbibliography

@book{jayaratna1994understanding,
  title={Understanding and evaluating methodologies: NIMSAD, a systematic framework},
  author={Jayaratna, Nimal},
  year={1994},
  publisher={McGraw-Hill, Inc.}
}

@book{eva1994ssadm,
  title={SSADM Version 4: A user's guide},
  author={Eva, Malcom},
  year={1994},
  publisher={McGraw-Hill, Inc.}
}

@incollection{beynon1998object,
  title={Object-Oriented Methods},
  author={Beynon-Davies, Paul},
  booktitle={Information Systems Development},
  pages={291--296},
  year={1998},
  publisher={Springer}
}

@inproceedings{cheatham1991object,
  title={Object-oriented vs. waterfall software development},
  author={Cheatham, Thomas J and Crenshaw, John H},
  booktitle={Proceedings of the 19th annual conference on Computer Science},
  pages={595--599},
  year={1991}
}

@article{charette2001decision,
  title={The decision is in: agile versus heavy methodologies. e-Project Management Advisory Service, Cutter Consortium, 2},
  author={Charette, R},
  year={2001}
}

@book{cobb2015project,
  title={The project manager's guide to mastering Agile: Principles and practices for an adaptive approach},
  author={Cobb, Charles G},
  year={2015},
  publisher={John Wiley \& Sons}
}

@book{schwaber2002agile,
  title={Agile software development with Scrum},
  author={Schwaber, Ken and Beedle, Mike},
  volume={1},
  year={2002},
  publisher={Prentice Hall Upper Saddle River}
}

@inproceedings{newkirk2002introduction,
  title={Introduction to agile processes and extreme programming},
  author={Newkirk, James},
  booktitle={Proceedings of the 24th International Conference on Software Engineering. ICSE 2002},
  pages={695--696},
  year={2002},
  organization={IEEE}
}

@book{collier2012agile,
  title={Agile analytics: A value-driven approach to business intelligence and data warehousing},
  author={Collier, Ken},
  year={2012},
  publisher={Addison-Wesley}
}

@article{beck2001manifesto,
  title={Manifesto for agile software development},
  author={Beck, Kent and Beedle, Mike and Van Bennekum, Arie and Cockburn, Alistair and Cunningham, Ward and Fowler, Martin and Grenning, James and Highsmith, Jim and Hunt, Andrew and Jeffries, Ron and others},
  year={2001},
  publisher={Snowbird, UT}
}

@inproceedings{anguelov2019research,
  title={Research for Usefulness of Agile Methods in Creative Business},
  author={Anguelov, Kiril},
  booktitle={2019 International Conference on Creative Business for Smart and Sustainable Growth (CREBUS)},
  pages={1--5},
  year={2019},
  organization={IEEE}
}

@article{sutherland2011scrum,
  title={The scrum papers: nut, bolts, and origins of an Agile framework},
  author={Sutherland, Jeff and Schwaber, Ken},
  journal={Scrum inc},
  year={2011}
}

@article{Schwaber2017,
  title={The Scrum GuideTM},
  author={K. Schwaber and J. Sutherland},
  journal={Scrum.org},
  year={2017}
}

@article{bhavsar2020scrum,
  title={Scrum: An agile process reengineering in software engineering},
  author={Bhavsar, Krunal and Shah, Vrutik and Gopalan, Samir},
  journal={International Journal of Innovative Technology and Exploring Engineering (IJITEE)},
  volume={9},
  number={3},
  pages={840--848},
  year={2020}
}

@article{ns2000scrum,
  title={The Scrum software development process for small teams},
  author={NS, Rising L Janoff},
  journal={IEEE Software},
  volume={174},
  pages={26--32},
  year={2000}
}

@inproceedings{cardozo2010scrum,
  title={SCRUM and productivity in software projects: a systematic literature review},
  author={Cardozo, Eliza SF and Ara{\'u}jo Neto, J Benito F and Barza, Alexandre and Fran{\c{c}}a, A C{\'e}sar C and da Silva, Fabio QB},
  booktitle={14th International Conference on Evaluation and Assessment in Software Engineering (EASE)},
  pages={1--4},
  year={2010}
}

@article{flora2014adopting,
  title={Adopting an agile approach for the development of mobile applications},
  author={Flora, Harleen K and Chande, Swati V and Wang, Xiaofeng},
  journal={International Journal of Computer Applications},
  volume={94},
  number={17},
  year={2014},
  publisher={Foundation of Computer Science}
}

@article{saleh2017comparative,
  title={Comparative study on the software methodologies for effective software development},
  author={Saleh, Sabbir M and Rahman, AM and Asgor, K Ali},
  journal={International Journal of Scientific \& Engineering Research},
  volume={8},
  number={4},
  pages={1018--1025},
  year={2017}
}

@book{beck2000extreme,
  title={Extreme programming explained: embrace change},
  author={Beck, Kent},
  year={2000},
  publisher={addison-wesley professional}
}

@article{abrahamsson2017agile,
  title={Agile software development methods: Review and analysis},
  author={Abrahamsson, Pekka and Salo, Outi and Ronkainen, Jussi and Warsta, Juhani},
  journal={arXiv preprint arXiv:1709.08439},
  year={2017}
}

@article{anwer2017comparative,
  title={Comparative analysis of two popular agile process models: extreme programming and scrum},
  author={Anwer, Faiza and Aftab, Shabib and Shah, SS Muhammad and Waheed, Usman},
  journal={International Journal of Computer Science and Telecommunications},
  volume={8},
  number={2},
  pages={1--7},
  year={2017}
}

@inproceedings{abrahamsson2003new,
  title={New directions on agile methods: a comparative analysis},
  author={Abrahamsson, Pekka and Warsta, Juhani and Siponen, Mikko T and Ronkainen, Jussi},
  booktitle={25th International Conference on Software Engineering, 2003. Proceedings.},
  pages={244--254},
  year={2003},
  organization={Ieee}
}

@online{WinNT,
  author = {Joshua Partogi},
  title = {Scrum And eXtreme Programming (XP)},
  year = 2018,
  url = {https://www.scrum.org/},
}

@inproceedings{srivastava2017scrum,
  title={SCRUM model for agile methodology},
  author={Srivastava, Apoorva and Bhardwaj, Sukriti and Saraswat, Shipra},
  booktitle={2017 International Conference on Computing, Communication and Automation (ICCCA)},
  pages={864--869},
  year={2017},
  organization={IEEE}
}

@inproceedings{saleh2019comparative,
  title={Comparative study within Scrum, Kanban, XP focused on their practices},
  author={Saleh, Sabbir M and Huq, Syed Maruful and Rahman, M Ashikur},
  booktitle={2019 International Conference on Electrical, Computer and Communication Engineering (ECCE)},
  pages={1--6},
  year={2019},
  organization={IEEE}
}

@inproceedings{taibi2017comparing,
  title={Comparing communication effort within the scrum, scrum with kanban, xp, and banana development processes},
  author={Taibi, Davide and Lenarduzzi, Valentina and Ahmad, Muhammad Ovais and Liukkunen, Kari},
  booktitle={Proceedings of the 21st International Conference on Evaluation and Assessment in Software Engineering},
  pages={258--263},
  year={2017}
}

@article{merzouk2017comparative,
  title={A Comparative Study of Agile Methods: Towards a New Model-based Method.},
  author={Merzouk, Soukaina and Elhadi, Sakina and Ennaji, Hassan and Marzak, Abdelaziz and Sael, Nawal},
  journal={Int. J. Web Appl.},
  volume={9},
  number={4},
  pages={121--128},
  year={2017}
}

@inproceedings{herdika2020variability,
  title={Variability and commonality requirement specification on agile software development: Scrum, xp, lean, and kanban},
  author={Herdika, Hana Rizky and Budiardjo, Eko K},
  booktitle={2020 3rd International Conference on Computer and Informatics Engineering (IC2IE)},
  pages={323--329},
  year={2020},
  organization={IEEE}
}

@article{merzouk2018agile,
  title={Agile Software Development: Comparative Study},
  author={Merzouk, Soukaina and Elhadi, Sakina and Cherkaoui, Abdessamad and Marzak, Abdelaziz and Sael, Nawal},
  journal={Smart Application and Data Analysis for Smart Cities (SADASC'18)},
  year={2018}
}

@article{anwer2018empirical,
  title={Empirical comparison of XP \& SXP},
  author={Anwer, Faiza and Aftab, Shabib and Bashir, Muhammad Salman and Nawaz, Zahid and Anwar, Madiha and Ahmad, Munir},
  journal={IJCSNS},
  volume={18},
  number={3},
  pages={161},
  year={2018}
}

@inproceedings{raza2018managing,
  title={Managing change in Agile software development a comparative study},
  author={Raza, Samrina and Waheed, Usman},
  booktitle={2018 IEEE 21st International Multi-Topic Conference (INMIC)},
  pages={1--8},
  year={2018},
  organization={IEEE}
}

@article{hosseini2020precise,
  title={Precise gamma based two-phase flow meter using frequency feature extraction and only one detector},
  author={Hosseini, S and Roshani, GH and Setayeshi, S},
  journal={Flow Measurement and Instrumentation},
  volume={72},
  pages={101693},
  year={2020},
  publisher={Elsevier}
}

@article{hosseini2021application,
  title={Application of Wavelet Feature Extraction and Artificial Neural Networks for Improving the Performance of Gas--Liquid Two-Phase Flow Meters Used in Oil and Petrochemical Industries},
  author={Hosseini, Siavash and Taylan, Osman and Abusurrah, Mona and Akilan, Thangarajah and Nazemi, Ehsan and Eftekhari-Zadeh, Ehsan and Bano, Farheen and Roshani, Gholam Hossein},
  journal={Polymers},
  volume={13},
  number={21},
  pages={3647},
  year={2021},
  publisher={MDPI}
}

@article{hosseini2022accurate,
  title={Accurate Flow Regime Classification and Void Fraction Measurement in Two-Phase Flowmeters Using Frequency-Domain Feature Extraction and Neural Networks},
  author={Hosseini, Siavash and Iliyasu, Abdullah M and Akilan, Thangarajah and Salama, Ahmed S and Eftekhari-Zadeh, Ehsan and Hirota, Kaoru},
  journal={Separations},
  volume={9},
  number={7},
  pages={160},
  year={2022},
  publisher={Multidisciplinary Digital Publishing Institute}
}

@article{bahiraei2020neural,
  title={Neural network modeling of thermo-hydraulic attributes and entropy generation of an ecofriendly nanofluid flow inside tubes equipped with novel rotary coaxial double-twisted tape},
  author={Bahiraei, Mehdi and Mazaheri, Nima and Hosseini, Siavash},
  journal={Powder technology},
  volume={369},
  pages={162--175},
  year={2020},
  publisher={Elsevier}
}

@article{bahiraei2021predicting,
  title={Predicting heat transfer rate of a ribbed triple-tube heat exchanger working with nanofluid using neural network enhanced by advanced optimization algorithms},
  author={Bahiraei, Mehdi and Foong, Loke Kok and Hosseini, Siavash and Mazaheri, Nima},
  journal={Powder Technology},
  volume={381},
  pages={459--476},
  year={2021},
  publisher={Elsevier}
}

@article{bahiraei2021neural,
  title={Neural network combined with nature-inspired algorithms to estimate overall heat transfer coefficient of a ribbed triple-tube heat exchanger operating with a hybrid nanofluid},
  author={Bahiraei, Mehdi and Foong, Loke Kok and Hosseini, Siavash and Mazaheri, Nima},
  journal={Measurement},
  volume={174},
  pages={108967},
  year={2021},
  publisher={Elsevier}
}

@article {hosseini2021increasing,
author = {Hosseini, Siavash and Setayeshi, Saeed and Roshani, Gholamhossein and Zahedi, Abdolhamid and Shama, Farzin},
title = {Increasing efficiency of two-phase flowmeters using frequency-domain feature extraction and neural network in the detector output spectrum},
journal = {Journal of Modeling in Engineering},
volume = {19},
number = {67},
pages = {47-57},
year  = {2021},
publisher = {semnan university},
issn = {2008-4854}, 
eissn = {2783-2538}, 
doi = {10.22075/jme.2021.19817.1860}
}

@INPROCEEDINGS{9918222,  author={Hosseini, Siavash and Shahbandegan, Amirmohammad and Akilan, Thangarajah},  booktitle={2022 IEEE Canadian Conference on Electrical and Computer Engineering (CCECE)},   title={Deep Neural Network Modeling for Accurate Electric Motor Temperature Prediction},   year={2022},  volume={},  number={},  pages={170-175},  doi={10.1109/CCECE49351.2022.9918222}}

@phdthesis{hosseini2023advanced,
  title={Advanced deep regression models for smart operation of the oil and gas industry},
  author={Hosseini, Siavash},
  year={2023}
}

@article{hosseini2023advancedt,
  title={Advanced Deep Regression Models for Forecasting Time Series Oil Production},
  author={Hosseini, Siavash and Akilan, Thangarajah},
  journal={arXiv preprint arXiv:2308.16105},
  year={2023}
}
\end{document}